\documentclass[structabstract]{aa}
\usepackage{graphicx}
\usepackage{txfonts}
 \usepackage{natbib}
\begin{document}
   \title{Physical parameters of close binaries QX And, RW Com, MR Del and BD +07$^{o}$ 3142}

   \author{G. Djura\v sevi\'c\inst{1,2},
	M. Y{\i}lmaz\inst{3},
 	\"{O}. Ba\c{s}t\"{u}rk\inst{4},
	T. K{\i}l{\i}\c{c}o\u{g}lu\inst{3},	
	O. Latkovi\'c\inst{1},
	\c{S}. \c{C}al{\i}\c{s}kan\inst{3}}

   \offprints{ G. Djurasevi\'c \vskip.2cm
   \noindent$^*$The data is available in electronic form at the CDS, via anonymous 
   ftp at cdsarc.u-strasbg.fr or via http://cdsweb.u-strasbg.fr/Abstract.html}

   \institute{Astronomical Observatory, Volgina 7, 11000 Belgrade, Serbia\\
	\email{gdjurasevic@aob.rs}
	\and Isaac Newton Institute of Chile, Yugoslavia Branch
	\and Ankara University, Faculty of Science, Dept. of Astronomy \& Space Sciences, TR-06100 Tandogan, Ankara - Turkey
	\and Ankara University, Astronomy \& Space Sciences Research and Application Center, TR-06837 Ahlatl{\i}bel, Ankara - Turkey\\}
	
   \titlerunning{QX And, RW Com, MR Del and BD +07$^{o}$ 3142}
   \authorrunning{G. Djurasevic, M. Y{\i}lmaz, \"{O}. Ba\c{s}t\"{u}rk et al.}

   \date{Received September 15, 1996; accepted March 16, 1997}


  \abstract
   {}
   {We analyze new multicolor light curves and recently published radial velocity curves for close binaries QX And, RW Com, MR Del, and BD +07$^{o}$ 3142 to determine the physical parameters of the components.}
   {The light curves are analyzed using a binary star model based on Roche geometry to fit the photometric observations. Spectroscopic parameters, such as the mass ratios and spectral types, were taken from recent spectroscopic studies of the systems in question. }
   {Our findings provide consistent and reliable sets of stellar parameters for the four studied binary systems. Of particular interest is the BD +07$^{o}$ 3142 system, since this is the first analysis of its light curves. We find that it is an overcontact binary of W UMa type and W subtype, and that each component has a large cool spot in the polar region. QX And is an A subtype, and RW Com a W subtype W UMa binary, and in both systems we find a bright spot in the neck region between the components. MR Del is a detached binary with a complex light curve that we could model with two cool spots on the hotter component. }
   {}


   \maketitle
%

\setcitestyle{authoryear,round,semicolon,aysep={},yysep={;}}

\section{Introduction}

 In this study we present simultaneous solutions of new, high-quality CCD light curves$^*$ and recently published radial velocity curves of four close binaries, QX And, RW Com, MR Del, and BD +07$^{o}$ 3142. While one of the systems, MR Del, is a close, detached system with a third light contribution from its visual companion, others have parameters consistent with the characteristics of the W UMa type systems. QX And belongs to the A subtype and RW Com to the W subtype. The analysis for BD +07$^{o}$ 3142, published here for the first time, shows that it is a W subtype W UMa system.

The information about observations and a general description of the light-curve analysis are given first, followed by the detailed history of previous study, the results of this work, and discussion for each star.


\section{Observations}

CCD photometric observations of the close binaries QX And, RW Com, MR Del, and BD +07$^{o}$ 3142 were obtained with the Apogee ALTA U47 CCD camera attached to the 40 cm Schmidt-Cassegrain telescope located at the Ankara University Observatory (AUG) by using BVR filters in close accordance with the Johnson-Cousins standard system. After calibrating  all the images (bias, dark, and flat correction) and using the IRAF task CCDPROC, we performed aperture photometry in the usual manner with the relevant tasks in the aperture photometry package (apphot) in IRAF\footnote{IRAF is distributed by the National Optical Astronomy Observatories, which are operated by the Association of Universities for Research in Astronomy, Inc.,  under cooperative agreement with the National Science Foundation.}. Both BVR magnitudes and their errors were computed as variable minus comparison. The nightly extinction coefficients for BVR magnitudes were determined from the observations of comparison stars. Orbital phases of the light curves were computed based on the light elements calculated using the times of minima recently published in the literature.

The log of the observations, information related to the comparison stars and the check stars used in the observations, and the light elements used for the calculation of photometric phases of the light curves are given in Tables 1, 2 and 3, respectively. All the the times of minima covered by our observations were calculated using the method of \citet{kwandvw56}, based on unweighted BVR observations. Times of minima being published for the first time are listed in Table 4. For RW Com, we used times of minima calculated by \citet{yilmaz09}.


\begin{table}
	\scriptsize
	\begin{flushleft}
	\begin{minipage}{140mm}
	\caption{The observation log.}
	\begin{tabular}{llcccccc}
	\noalign{\smallskip}
	\hline
	\noalign{\smallskip}
	System & Obs. date &  \multicolumn{6}{c}{Nightly mean errors [mag] and integration times [sec]} \\
	\noalign{\smallskip}
	\hline
	\noalign{\smallskip}
	& & \multicolumn{2}{c}{B} & \multicolumn{2}{c}{V} & \multicolumn{2}{c}{R} \\
	\noalign{\smallskip}
	\hline
	\noalign{\smallskip}
	QX And 
		& 2009/08/17 & $\pm$0.005 & 15  & $\pm$0.004 & 5 & $\pm$0.004 & 3\\
		& 2009/09/25 & $\pm$0.005 & 15  & $\pm$0.004 & 5 & $\pm$0.004 & 3\\
		& 2009/09/29 & $\pm$0.005 & 15  & $\pm$0.004 & 5 & $\pm$0.004 & 3\\
	
	RW Com 
		& 2009/02/02 & $\pm$0.010 & 40 & $\pm$0.005 & 17 & $\pm$0.004 & 10\\
		& 2009/03/20 & $\pm$0.010 & 40 & $\pm$0.005 & 18 & $\pm$0.004 & 10\\
		& 2009/03/23 & $\pm$0.010 & 35 & $\pm$0.005 & 20 & $\pm$0.004 & 10\\
		& 2009/04/08 & $\pm$0.010 & 40 & $\pm$0.005 & 17 & $\pm$0.004 & 10\\
	 
	MR Del 
		& 2009/06/16 & $\pm$0.003 & 10 & $\pm$0.003 & 3 & $\pm$0.003 & 2\\
		& 2009/07/29 & $\pm$0.003 & 25 & $\pm$0.003 & 4 & $\pm$0.003 & 2\\
		& 2009/08/15 & $\pm$0.003 & 25 & $\pm$0.003 & 4 & $\pm$0.003 & 2\\
		& 2009/09/25 & $\pm$0.003 & 20 & $\pm$0.003 & 8 & $\pm$0.003 & 3\\

	BD
		& 2009/08/07 & $\pm$0.003 & 30 & $\pm$0.002 & 9 & $\pm$0.002 & 5\\
	+07$^{o}$ 3142	
		& 2009/08/09 & $\pm$0.003 & 30 & $\pm$0.002 & 9 & $\pm$0.002 & 5\\
		& 2009/08/19 & $\pm$0.003 & 30 & $\pm$0.002 & 10 & $\pm$0.002 & 5\\
		& 2009/08/31 & $\pm$0.003 & 35 & $\pm$0.002 & 10 & $\pm$0.002 & 6\\
		& 2009/09/01 & $\pm$0.003 & 35 & $\pm$0.002 & 10 & $\pm$0.002 & 6\\	
	\noalign{\smallskip}
	\hline
	\noalign{\smallskip}
	\end{tabular}
	\end{minipage}
	\end{flushleft}
\end{table}


\begin{table}
	\scriptsize
	\begin{flushleft}
	\begin{minipage}{140mm}
	\caption{The light elements used in this study for each system.}
	\begin{tabular}{lccc}
	\noalign{\smallskip}
	\hline
	\noalign{\smallskip}
	& Epoch & & \\
	System & (HJD+2400000) & Period (days) & Ref. \\
	\noalign{\smallskip}
	\hline
	\noalign{\smallskip}
	QX And & 55101.4475(19) & 0.4121716(9)  & This study \\
	RW Com & 54272.8826(6) & 0.2373463 (1)  & This study \\
	MR Del & 55042.5055(7) & 0.5216903(1)   & This study  \\
	BD +07$^{o}$ 3142 & 54688.3631(12) & 0.275205(1)  & This study  \\
	\noalign{\smallskip}
	\hline
	\noalign{\smallskip}
	\end{tabular}
	\end{minipage}
	\end{flushleft}
\end{table}


 \begin{table} [ht]
	\scriptsize
	\begin{flushleft}
	\begin{minipage}{140mm}
	\caption{The comparison and check stars used during the observation of each close binary.}
	\begin{tabular}{llll}
	\noalign{\smallskip}
	\hline
	\noalign{\smallskip}
	& Variable  & Comparison & Check  \\
	\noalign{\smallskip}
	\hline
	\noalign{\smallskip}
	& QX And & GSC 2816-2039 & GSC 2816-1820 \\
	$\alpha_{2000}$: & 01$^{h}$57$^{m}$57$^{s}$.78 & 01$^{h}$57$^{m}$47$^{s}$.14 & 01$^{h}$57$^{m}$31$^{s}$.94   \\
	$\delta_{2000}$: & +37$^{\circ}$48$^{'}$22$^{"}$.4 & +37$^{\circ}$47$^{'}$30$^{"}$.3 & +37$^{\circ}$53$^{'}$43$^{"}$.2 \\
	Spec. Type: & F5 & F0 & F3 \\
	V(mag): & 11.43 & 10.9 & 11.2 \\
	\noalign{\smallskip}
	\hline
	\noalign{\smallskip}
	& RW Com & GSC 1991-1657 & - \\
	$\alpha_{2000}$: & 12$^{h}$33$^{m}$00$^{s}$.28 & 12$^{h}$33$^{m}$03$^{s}$.94 & - \\
	$\delta_{2000}$: & +26$^{\circ}$42$^{'}$58$^{"}$.4 & +26$^{\circ}$37$^{'}$27$^{"}$.6 & - \\
	Spec. Typ.: & G2 & - & - \\
	V(mag): & 11.00 & 11.86 & - \\
	\noalign{\smallskip}
	\hline
	\noalign{\smallskip}
	& MR Del & GSC 518-13 & GSC 518-447 \\
	$\alpha_{2000}$: & 20$^{h}$31$^{m}$13$^{s}$.46 & 20$^{h}$31$^{m}$24$^{s}$.18 & 20$^{h}$31$^{m}$23$^{s}$.53 \\
	$\delta_{2000}$: & +05$^{\circ}$13$^{'}$08$^{"}$.5  & +05$^{\circ}$12$^{'}$50$^{"}$.1 & +05$^{\circ}$18$^{'}$03$^{"}$.1 \\
	Spec. Type: & K0 & - & F2 \\
	V(mag): & 11.01 & 9.86 & 9.73 \\
	\noalign{\smallskip}
	\hline
	\noalign{\smallskip}
	& BD +07$^{o}$ 3142 & GSC 380-167 & GSC 380-429 \\
	$\alpha_{2000}$: & 16$^{h}$10$^{m}$03$^{s}$.21 & 16$^{h}$20$^{m}$09$^{s}$.22 & 16$^{h}$20$^{m}$12$^{s}$.84 \\
	$\delta_{2000}$: & +07$^{\circ}$07$^{'}$28$^{"}$.7  & +07$^{\circ}$03$^{'}$54$^{"}$.6 & +07$^{\circ}$11$^{'}$17$^{"}$.5 \\
	Spec. Typ.: & G5 & F8 & - \\
	V(mag): & 9.89 & 9.15 & 10.6 \\
	\noalign{\smallskip}
	\hline
	\noalign{\smallskip}
	\end{tabular}
	\end{minipage}
	\end{flushleft}
\end{table}


\begin{table} [ht]
	\scriptsize
	\begin{flushleft}
	\begin{minipage}{140mm}
	\caption{New times of minima derived from our BVR observations.}
	\begin{tabular}{lll}
	\noalign{\smallskip}
	\hline
	\noalign{\smallskip}
	Variable  & Times of min. & Type  \\
	\noalign{\smallskip}
	\hline
	\noalign{\smallskip}
	QX And & 2455061.4672(1) & I \\
	& 2455101.4475(2) & I \\
	& 2455104.3342(2) & I \\
	& 2455104.5398(1) & II \\
	\noalign{\smallskip}
	\hline
	\noalign{\smallskip}
	MR Del & 2455042.5056(1) & I \\
	& 2454999.4669(1) & II \\
	\noalign{\smallskip}
	\hline
	\noalign{\smallskip}
	BD +07$^{o}$ 3142 & 2454688.3643(2) & I \\
	& 2454686.2978(2) & II \\
        \noalign{\smallskip}
	\hline
	\noalign{\smallskip}
	\end{tabular}
	\end{minipage}
	\end{flushleft}
\end{table}


\section{Light curve analysis}
\label{analysis}

We analyzed all the light curves using the program by \citet{djur92a}, generalized for the case of an overcontact configuration \citep{djur98}. The program is based on the Roche model and the principles described by \citet{wilanddev71}. The system parameters are estimated by applying an inverse-problem method \citep{djur92b}, based on the modified \citet{marquardt} algorithm. The underlying binary system model allows various system configurations from detached to overcontact, including active spot regions on the components. The latest version of the model and the solving procedure is described in detail by \citet{djura08}.

The fitting was done for a limited set of parameters, while the following parameters were fixed based on information from independent analysis of the radial velocity curves and plausible assumptions about the systems.

\begin{itemize}
\item{The mass ratio $q=m_c/m_h$ (where the subscripts $h$ and $c$ stand for the hotter and cooler component) for each system was fixed based on the radial velocity studies by \citet{pribb09} and \citet{ruca08}.}
\item{The  temperature of the hotter component $T_h$ for each system was fixed according to its spectral type and the revised theoretical ${\rm T_{eff}}$-spectral type calibration by \citet{mart05}.}
\item{The gravity-darkening coefficients $\beta_{h,c}$ and the albedos $A_{h,c}$ for each system were fixed to their theoretical values according to the the temperature of the hotter component and a preliminary estimate of the temperature of the cooler component.}
\item{The ratio of the rotation rate to the Keplerian orbital rate $f_{h,c}=\omega_{h,c}/ \omega_K$ for all systems was fixed in accordance with the assumption that the rotation of the components is synchronous with the orbital revolution since the tidal effect are expected to contribute to the synchronization of the rotational and orbital periods.}
\end{itemize}

The limb darkening in the latest version of the model follows the nonlinear approximation given by \citet{claret00} and the limb-darkening coefficients for the corresponding passbands were interpolated from Claret's tables according to the current values of ${\rm T_{eff}}$ and $\log(g)$ in each iteration.

The results of the analysis are given for each star in a table (see Tables~\ref{TabQXAnd},~\ref{TabRWCom},~\ref{TabMRDel}, and~\ref{TabBD}), where $n$ is the total number of the B, V, and R observations; ${\rm \Sigma (O-C)^2}$ - the final sum of squares of residuals between observed (LCO) and synthetic (LCC) light-curves; ${\rm \sigma}$ - the standard deviation of the residuals; $q=m{\rm _c}/m{\rm _h}$ - the mass ratio of the components; ${\rm T_{\rm h,c}}$ - the temperatures of the hotter and the cooler component; $\beta_{\rm h,c}$, ${\rm A_{\rm h,c}}$, $f_{\rm h,c}$  - the gravity-darkening exponents, albedos and nonsynchronous rotation coefficients of the components; ${\rm A_{s}}$, ${\rm \theta_{s}}$, ${\rm \lambda_{s}}$, and ${\rm \varphi_{s}}$ - the temperature coefficients, angular dimensions, longitudes, and latitudes (in arc degrees) of the spots (if present in the model); ${\rm F_{h,c}}$ - the filling factors for the critical Roche lobe of the hotter and cooler component; $i \ [^\circ]$ - the orbit inclination (in arc degrees); ${\rm L_3/(L_h+L_c+L_3)}$ - the third light contribution; $\Omega_{\rm h,c}$, $\Omega_{in}$ and $\Omega_{out}$ - the dimensionless surface potentials of the system components and of the inner and outer contact surfaces; $f{\rm _{over}}[\%]$ - the degree of overcontact; ${\rm R_{h,c}}$ - the polar radii of the components in units of separation; ${\rm L_h/(L_h+L_c)}$ - the  luminosity {\rm (B;V;R)} of the hotter star; $\cal M_{\rm h,c} {\rm [M_{\odot}]}$, $\cal R_{\rm h,c} {\rm [R_{\odot}]}$ - the stellar masses and mean radii of the components in solar units; ${\rm log} \ g_{\rm h,c}$ - the logarithm (base 10) of the system components effective gravity; $M^{\rm h,c}_{\rm bol}$ - the absolute bolometric magnitudes of the components; and $a_{\rm orb} {\rm [R_{\odot}]}$ - the orbital semi-major axis in units of solar radius.

Although the mass ratio for each system was fixed when solving the inverse problem to the value adopted from the relevant spectroscopic study, the influence of the mass ratio uncertainty was incorporated in estimating the final parameter errors. Namely, we made trial runs by fixing the mass ratio to its highest and lowest possible values (chosen according to the reported uncertainty), and the deviations of parameter values obtained in this way were combined with the formal fitting errors (in the sense of adopting the highest error). Tables~\ref{TabQXAnd},~\ref{TabRWCom},~\ref{TabMRDel}, and~\ref{TabBD} list the parameter uncertainties estimated in this way. Still, the real parameter uncertainties may be greater than the tabulated errors since a significant contribution to the errors may come from the uncertainty of the effective temperature of the hotter star, which was in each case fixed on the basis of its spectral type.

The distances to the four systems were computed based on the apparent magnitudes taken from the SIMBAD database and the computed absolute magnitudes, with corrections for the interstellar absorption. We used the bolometric corrections from the tabulation of \citet{flower96}, according to the computed effective temperatures of the components for each system. The interstellar extinction values ($A_v$) in the V passband were computed using the reddening values estimated from the infrared dust emission maps of \citet{schleg98} and by assuming an extinction to reddening ratio of  $A_v/E(B-V)$ of 3.1. \citet{schleg98} refer to the total absorption.

In the following sections we give the detailed history of previous knowledge and the results of the present study for each system in turn.


\section{QX And}

QX And (GSC 02816-01950, NGC 752-235) is a short-period eclipsing binary, located in the intermediate-age open cluster NGC 752 (OC1 363 in \citealt{rupr88}, age=$2.0 \pm 0.2 \times 10^9$ yr from \citealt{dan94}). The system is also an x-ray source \citep{belver96}. The light variability of the system was first noted by \citet{john53}, and \citet{twa83} pointed out that QX And was probably a binary based on its location in the color-magnitude diagram. 


\begin{figure*}
\includegraphics{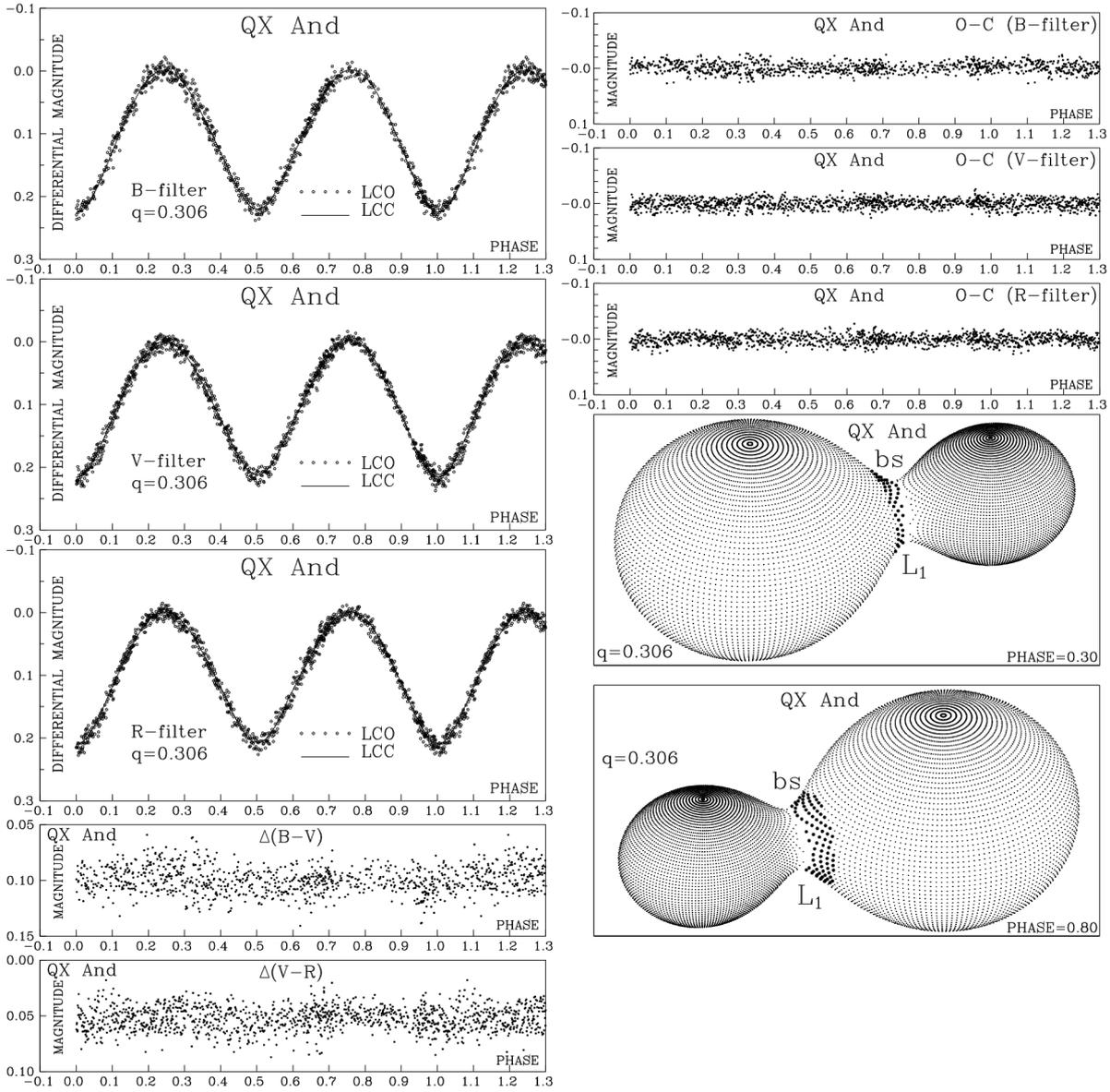}
\caption{Observed (LCO) and synthetic (LCC) light curves of {\rm QX And},
obtained by simultaneous analysis of the B, V, and R observations,
$\Delta(B-V)$, $\Delta(V-R)$ color curve, final O-C residuals, and the
view of the system model at the orbital phases 0.30 and 0.80, obtained
with parameters from the inverse problem solution.}
\label{fQXAnd}
\end{figure*}


The first photometric and spectroscopic observations of QX And were obtained by \cite{mila95}. They analyzed the light and radial velocity curves simultaneously with the Wilson-Devinney code \citep{wilanddev71}, and obtained a model of an A-type W UMa system with a contact parameter of $f = 0.21 \pm 0.11$ that fit the observations best. The authors estimated the spectral type of the system as F3-5, determined the absolute parameters of the components ($M_{1}=1.18 M_{\odot}$ and $M_{2}=0.24 M_{\odot}$), and found the distance to the binary and the cluster to be 381 $\pm$ 17 pc. The radial velocity data of QX And was derived using cross-correlation functions. Unfortunately, the RV observations of \citet{mila95} were of rather poor quality, resulting in only marginally useful orbital parameters. The authors obtained four light-curve solutions (the spotless, one-spot, two-spot, and three-spot cases) based on their finding that QX And displayed night-to-night variations in its light curves. Although the three-spot model fit the observations best, they derived the absolute parameters from the spotless-case solution. The authors assumed all spots were located on the hotter component.

\citet{mila95} located a spot on one of the components although they performed their solution under the assumption of a radiative envelope. Noticing the mistake in this approach, \citet{qiana07} find that QX And is a deep overcontact binary system with a high degree of overcontact (f = 55.9\%) and a low mass ratio (q = 0.2327). From the published times of minima, they point out that the orbital period is changing at a rate of $dp/dt=2.48\times 10^{-7}$ days/yr, which indicates a mass transfer from the less massive to the more massive component. Because the light curve they obtained displayed no asymmetry, they did not locate any spots on either of the components.


\begin{table}
\begin{flushleft}
\caption{Results of the simultaneous (BVR) analysis of the {\rm QX And}
light curves obtained by solving the inverse problem for the Roche model
with a bright spot on the more massive (hotter) component.}
\label{TabQXAnd}

\[
\begin{array}{ll}
\hline
\noalign{\smallskip}
{\rm Quantity} &  \\
\noalign{\smallskip}
\hline
\noalign{\smallskip}
n & 2689 \\
{\rm \Sigma(O-C)^2} & 0.1996 \\
{\rm \sigma} & 0.0086 \\
q=m_{\rm c}/m_{\rm h} & 0.306 \pm 0.009\\
T_{\rm h} & 6440 \\
A_{\rm c}=A_{\rm h} & 0.5 \\
\beta_{\rm c}=\beta_{\rm h}  & 0.08 \\
f_{\rm c}=f_{\rm h} & 1.0 \\
{\rm A_{hs}=T_{bs}/T_h} & 1.02 \pm 0.01 \\
{\rm \theta_{bs}} & 19.6 \pm 1.0 \\
{\rm \lambda_{bs}} &357.3 \pm 1.4 \\
{\rm \varphi_{bs}} & 0 \\
T_{\rm c} & 6420 \pm 20 \\
F_{\rm h} &1.031 \pm 0.001 \\
i \ [^\circ] &54.6  \pm 0.2 \\
\Omega_{\rm h,c} & 2.41 \pm 0.02 \\
\Omega_{\rm in},\Omega_{\rm out} & 2.48 \pm 0.02 \ , 2.29 \pm 0.02 \\
f_{\rm over} [\%] & 35.2 \pm 2.0\\
R_{\rm h} [D=1] & 0.47 \pm 0.01\\
R_{\rm c} [D=1] & 0.28 \pm 0.01\\
{\rm L_h/(L_h+L_c)} & 0.74 [{\rm B}] ;\ 0.74 [{\rm V}] ;\ 0.74 [{\rm R}] \\
\cal M_{\rm h} {\rm [M_{\odot}]}  & 1.47 \pm 0.05 \\
\cal M_{\rm c} {\rm [M_{\odot}]}  & 0.45 \pm 0.02 \\
\cal R_{\rm h} {\rm [R_{\odot}]}  & 1.46 \pm 0.02 \\
\cal R_{\rm c} {\rm [R_{\odot}]}  & 0.88 \pm 0.02 \\
{\rm log} \ g_{\rm h} & 4.27 \pm 0.03 \\
{\rm log} \ g_{\rm c} & 4.20 \pm 0.03 \\
M^{\rm h}_{\rm bol} & 3.49 \pm 0.03 \\
M^{\rm c}_{\rm bol} & 4.60 \pm 0.03 \\
a_{\rm orb} {\rm [R_{\odot}]} & 2.89 \pm 0.04 \\
d[\rm pc] & 416 \pm 46 \\
\noalign{\smallskip}
\hline
\end{array}
\]
\end{flushleft}
\end{table}

\citet{pribb09} obtained radial velocity curves of both components and determined a mass ratio of q = 0.306. The authors estimated the spectral type of the system as F8. This recent mass ratio is more reliable than the one (q = 0.203) found by \citet{mila95}. The difference in the mass ratios found by these two authors is substantial, which is why new analysis, based on a reliable, spectroscopically estimated mass ratio and quality photometric observations was required for a better understanding of this system.

The results of our photometric analysis, based on the spectroscopically estimated mass ratio  given by \citet{pribb09}, are presented in Table~\ref{TabQXAnd}. Figure~\ref{fQXAnd} shows the observed (LCO) and the synthetic (LCC) light curves in the B, V, and R filters (upper left), the $B-V$ and $V-R$ color indices (lower left), the $O-C$ residuals (upper right) and the geometrical model of the system in representative phases 0.3 and 0.8 (lower right). The parameters are given with the uncertainties estimated by combining the formal nonlinear, least-squared fitting errors with the errors arising from the uncertainty of the spectroscopic mass ratio ($q=0.306\pm 0.009$), as described in Section~\ref{analysis}.

Our solution shows that the less massive and slightly cooler component eclipses the more massive star in the primary minimum (at phase zero), placing QX And in the A subclass of W UMa type systems, with a significant overcontact degree of $f_{\rm over}\approx 35\%$, and a small temperature difference between the components. The distance estimation from the absolute magnitude computed with our model agrees with the estimation of $380\pm \rm36 pc$ obtained by \citet{bilir05}, using the parallax value.

We did not observe any night-to-night variations in the light curves. The absence of significant phase dependence of color curves in Figure~\ref{fQXAnd} indicates that there is good thermal contact between the components. However, for successful fitting of the slight asymmetry in the light curves, the optimal model of the system needs a bright spot located in the neck region of the more massive, hotter component. This result agrees with the analysis by \citet{qiana07}, who find that mass transfer from the less massive to the more massive component, indicated by their O-C analysis, could justify a possible hot spot in the neck region of the more massive component. With the large difference in the masses of components, their equal temperatures suggest a significant energy exchange through the neck region.

Because of the specific light curve shape with almost equal depths of the primary and secondary eclipses and a relatively small amplitude resulting from the low orbital inclination ($i\approx 55^\circ$), we also tested the hypothesis that the system is in a W subtype W UMa configuration. However, this model gave a significantly worst fit to the observations.

In comparison with previous studies of QX And by \citet{mila95} and \citet{qiana07}, who used the underestimated mass ratio of $\rm q \approx 0.2$ obtained from the radial velocities of \citet{mila95}, our solution yields a significantly more massive secondary (${\cal M}_{\rm c,\ Milone}=0.24\ M_\odot$, ${\cal M}_{\rm c,\ Quian}=0.27\ M_\odot$, and ${\cal M}_{\rm c,\ this\ study}=0.45\ M_\odot$), an overcontact parameter that is between their estimated values ($f_{\rm Milone}=21\%$, $f_{\rm Quian}=55\%$, and $f_{\rm this\ study}=35\%$), somewhat larger radii of the components (${\cal R}_{\rm c,\ Milone}=0.68\ R_\odot$, ${\cal R}_{\rm h,\ Milone}=1.39\ R_\odot$, and ${\cal R}_{\rm c,\ this\ study}=0.88\ R_\odot$, ${\cal R}_{\rm h,\ this\ study}=1.46\ R_\odot$), and a lower inclination ($i_{\rm Milone}\approx58^\circ$, $i_{\rm Quian}\approx56^\circ$, and $i_{\rm this\ study}\approx55^\circ$). The temperatures, absolute bolometric magnitudes, and the estimated distance of the system agree well between these two studies and our own.

\section{RW Com}

RW Com (HIP 61243) is a short-period eclipsing binary. Its variability was first recognized by \citet{jor23}, who discovered it while observing the Cepheid variable S Comae and classified it as $\beta$ Lyrae type binary. \citet{stra50} performed the first spectral observations of the system and suggested that the emission components in the Ca II absorption line, observed to be strong in both conjunctions and weak in quadratures, originated in both stars.


\begin{figure*}
\includegraphics{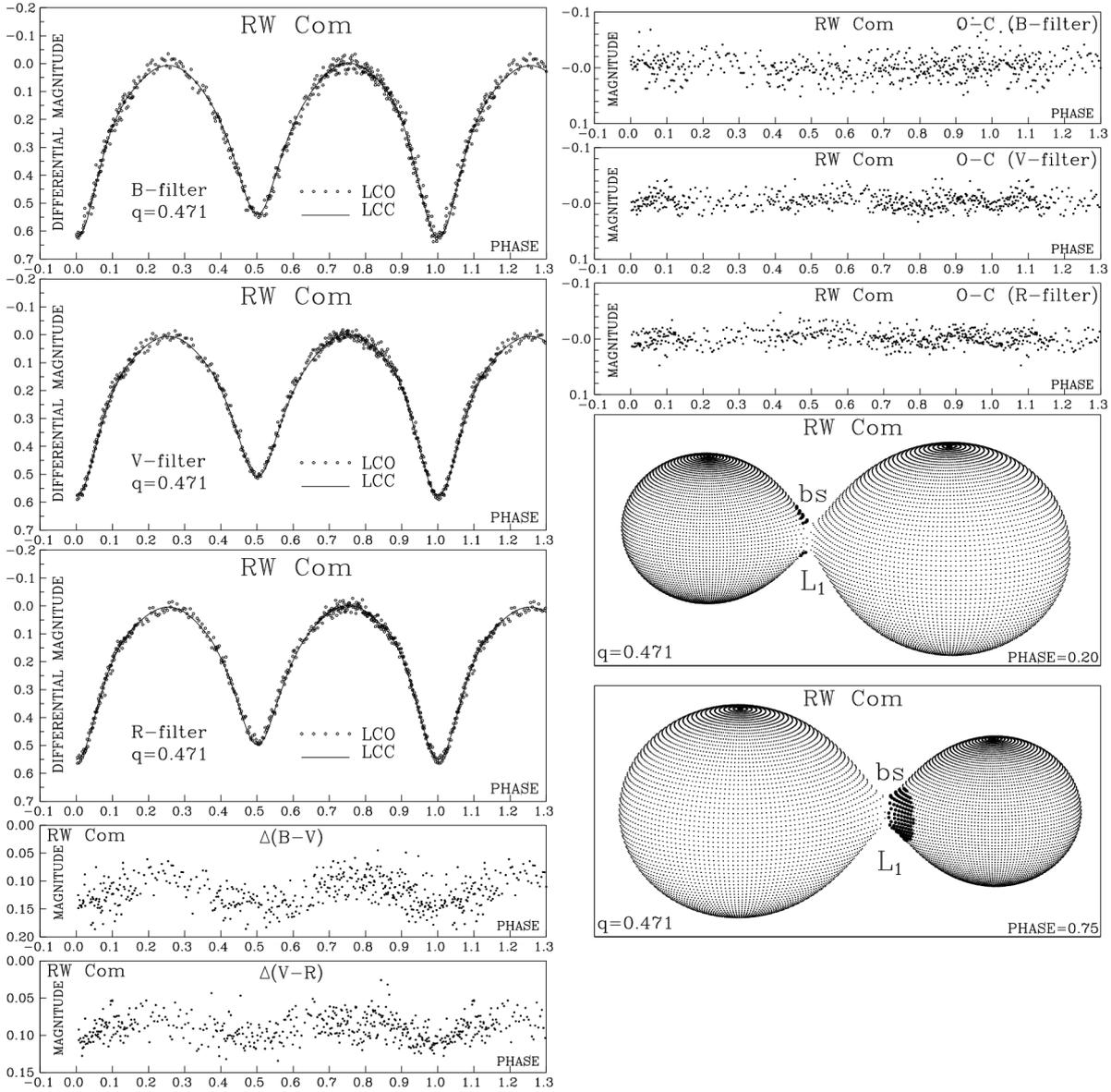}
\caption{Observed (LCO) and synthetic (LCC) light curves of {\rm RW Com},
obtained by simultaneous analysis of the B, V, and R observations,
$\Delta(B-V)$, $\Delta(V-R)$ color curve, final O-C residuals, and the
view of the system model at the orbital phases 0.20 and 0.75, obtained
with parameters from the inverse problem solution.}
\label{fRWCom}
\end{figure*}



\citet{oconn51} noted the asymmetry and variability of the light curves, and \citet{mila76} interpreted them as evidence of mass loss from the system, primarily through the outer Lagrangian point, facing the observer at primary minimum. \citet{mila80} constructed light curves based on all data up to the time of their study, and also noted the variability in the light curves and the differences between the light levels of maxima. The asymmetry was reported as increasing with wavelength by \citet{dav81}.


\begin{table}[ht]
\begin{flushleft}
\caption{Results of the simultaneous (BVR) analysis of the {\rm RW Com}
light curves obtained by solving the inverse problem for the Roche model
with a bright spot on the less massive (hotter) component.}
\label{TabRWCom}
\[
\begin{array}{ll}
\hline
\noalign{\smallskip}
{\rm Quantity} &  \\
\noalign{\smallskip}
\hline
\noalign{\smallskip}
n & 1309 \\
{\rm \Sigma(O-C)^2} & 0.3345 \\
{\rm \sigma} & 0.0160 \\
q=m_{\rm h}/m_{\rm c} & 0.471 \pm 0.006\\
T_{\rm h} & 4900 \\
A_{\rm c}=A_{\rm h} & 0.5 \\
\beta_{\rm c}=\beta_{\rm h}  & 0.08 \\
f_{\rm c}=f_{\rm h} & 1.0 \\
{\rm A_{bs}=T_{bs}/T_h} & 1.09 \pm 0.02 \\
{\rm \theta_{bs}} & 14.2 \pm 1.0 \\
{\rm \lambda_{bs}} &169.9 \pm 3.0 \\
{\rm \varphi_{bs}} &  0 \\
T_{\rm c} & 4720 \pm 20 \\
F_{\rm c} &1.007 \pm 0.001 \\
i \ [^\circ] & 74.9 \pm 0.1 \\
\Omega_{\rm c,h} & 2.80 \pm 0.01 \\
\Omega_{\rm in},\Omega_{\rm out} & 2.82 \pm 0.01 \ , 2.54 \pm 0.01 \\
f_{\rm over} [\%] & 6.1 \pm 1.0 \\
R_{\rm c} [D=1] & 0.42 \pm 0.01 \\
R_{\rm h} [D=1] & 0.30 \pm 0.01 \\
{\rm L_c/(L_c+L_h)} & 0.60 [{\rm B}] ;\ 0.61 [{\rm V}] ;\ 0.62 [{\rm R}] \\
\cal M_{\rm c} {\rm [M_{\odot}]}  & 0.80 \pm 0.02 \\
\cal M_{\rm h} {\rm [M_{\odot}]}  & 0.38 \pm 0.02 \\
\cal R_{\rm c} {\rm [R_{\odot}]}  & 0.77 \pm 0.02 \\
\cal R_{\rm h} {\rm [R_{\odot}]}  & 0.54 \pm 0.01 \\
{\rm log} \ g_{\rm c} & 4.57 \pm 0.03 \\
{\rm log} \ g_{\rm h} & 4.54 \pm 0.03 \\
M^{\rm c}_{\rm bol} & 6.24 \pm 0.05 \\
M^{\rm h}_{\rm bol} & 6.82 \pm 0.05 \\
a_{\rm orb} {\rm [R_{\odot}]} & 1.70\pm 0.06 \\
d[\rm pc] & 91 \pm 5 \\
\noalign{\smallskip}
\hline
\end{array}
\]
\end{flushleft}
\end{table}
From their spectroscopic study, \citet{mila85} found the mass ratio of 0.34 and classified the system as a W-subtype W UMa binary of spectral type G5-G8. Based on the results of this study, \citet{mila87} analyzed UBV light curves of the system with the Wilson-Devinney code. Fixing the temperature to three different values  (5400 K, 5600 K, and 5800 K) according to the estimated spectral type range (G5-G8), they tested different solutions, including both cold and hot photospheric spot configurations, and gave the absolute parameters of the solution that fit the observations best.

\citet{sriv87} published the first O-C curve of the system based on the previously published times of minima, and found a cyclic variation in the orbital period. Relying on the closely spaced data obtained between 1967 and 1986, \citet{sriv87} suggested that a third body with an orbital period of 16 years was causing the period to vary in time. \citet{qiana02}, however, found that the variation caused by the third body had a 13.3 year period, and was superimposed on a decreasing trend that the author explained by angular momentum loss from the system. \citet{yangli03} considered both mass exchange and mass loss mechanisms that could cause the observed decrease in the period but found that the cyclic variation superimposed on the decreasing trend was quasi-periodic; hence, it could only be attributed to magnetic activity.

\citet{pribruca06} and \citet{ruca07} did not detect the signature of a third body in their studies dedicated to detecting faint third bodies in contact binary systems. In addition, broadening functions constructed for radial velocity analysis by \citet{priba09} did not show any evidence of a third companion in the system. They also found a much higher mass ratio (0.471) than that found by \citet{mila85}, on which all the light curve analyses published up to that time were based. Moreover, \citet{priba09} determined a different spectral type (K2 V) for the system, and conclude that the observations \citet{mila85} used were subject to errors because of the heavy blends in lines used in their cross-correlation analysis, and the long exposure times they employed in the observations.

The results of our photometric analysis, based on spectroscopic elements estimated by \citet{priba09} are given in Table~\ref{TabRWCom}. Figure~\ref{fRWCom} shows the observed (LCO) and the synthetic (LCC) light curves in the B, V, and R filters (upper left), the $B-V$ and $V-B$ color indices (lower left), the $O-C$ residuals (upper right), and the geometrical model of the system in representative phases 0.20 and 0.75 (lower right). The tabulated parameter uncertainties were estimated by combining the formal nonlinear least-squared fitting errors with the errors arising from the uncertainty of the spectroscopic mass ratio ($q= 0.471 \pm 0.006$), as described in Section~\ref{analysis}.

Our solution shows that RW Com belongs to the W subclass of the W UMa type systems, i. e., the less massive, but hotter component, is eclipsed in the deeper primary minimum (at phase zero). The system is in a slightly overcontact configuration with $f_{\rm over}\approx 6\%$, and the temperature difference between the components is small ($\Delta {\rm T=T_h-T_c}\approx 180$ K). The optimal model of the system includes a bright spot in the neck region of the less massive, hotter component, enabling successful fitting of the slightly asymmetric light curves. This assumption is in accordance with mass exchange from the more massive primary to the secondary component that would cause the observed decrease in orbital period. At the same time, the hot region in the neck zone of the common envelope, located on the less massive secondary, can be interpreted as one consequence of an intensive energy transfer from the primary to the secondary.

The distance of the system, estimated from the absolute magnitude obtained with our model, is within the uncertainty limits of the distance based on the new Hipparcos parallax of $85 \pm 18 \rm pc$, given by \citet{vanlee07}. We found it unnecessary to include a third light in the solution, and that supports the findings of \citet{priba09}, who found no spectroscopic evidence of a third body in the system. Comparison of our results with the parameters reported by \citet{mila87}, which were based on the underestimated value of the mass ratio from \citet{mila85}, show that our solution gives higher masses of the components (${\cal M}_{\rm c,\ Milone}=0.56\ M_\odot$, ${\cal M}_{\rm h,\ Milone}=0.20\ M_\odot$ and ${\cal M}_{\rm c,\ this\ study}=0.80\ M_\odot$, ${\cal M}_{\rm h,\ this\ study}=0.38\ M_\odot$), a smaller overcontact factor ($f_{Milone}=17\%$ and $f_{this\ study}=6\%$) and larger radii (${\cal R}_{\rm c,\ Milone}=0.67\ R_\odot$, ${\cal R}_{\rm h,\ Milone}=0.41\ R_\odot$ and ${\cal R}_{\rm c,\ this\ study}=0.77\ R_\odot$, ${\cal R}_{\rm h,\ this\ study}=0.54\ R_\odot$), with approximately the same inclination and considerably lower temperatures, which is a consequence of the cooler spectral type assigned to this system by \citet{priba09}. However, the absolute bolometric magnitudes and the estimated distances are in relatively good agreement.


\section{MR Del}

MR Del (GSC 00518-1755, HD 195434, ADS 13940) is the brighter component of a visual binary system. It is an eclipsing binary \citep{cuti97}, separated from its visual companion by 1$^{"}$.8 at 71$^{\circ}$ \citep{mason01}. The whole system was classified as a halo population object or an old disk star by \cite{cuti97}, based on its proper motion and low metallicity \citep{carney94,sandkow86}. This makes it an interesting object, because it displays soft X-ray emission \citep{pye95} and photospheric spot activity \citep{cuti97} despite its age.


\begin{figure*}
\includegraphics{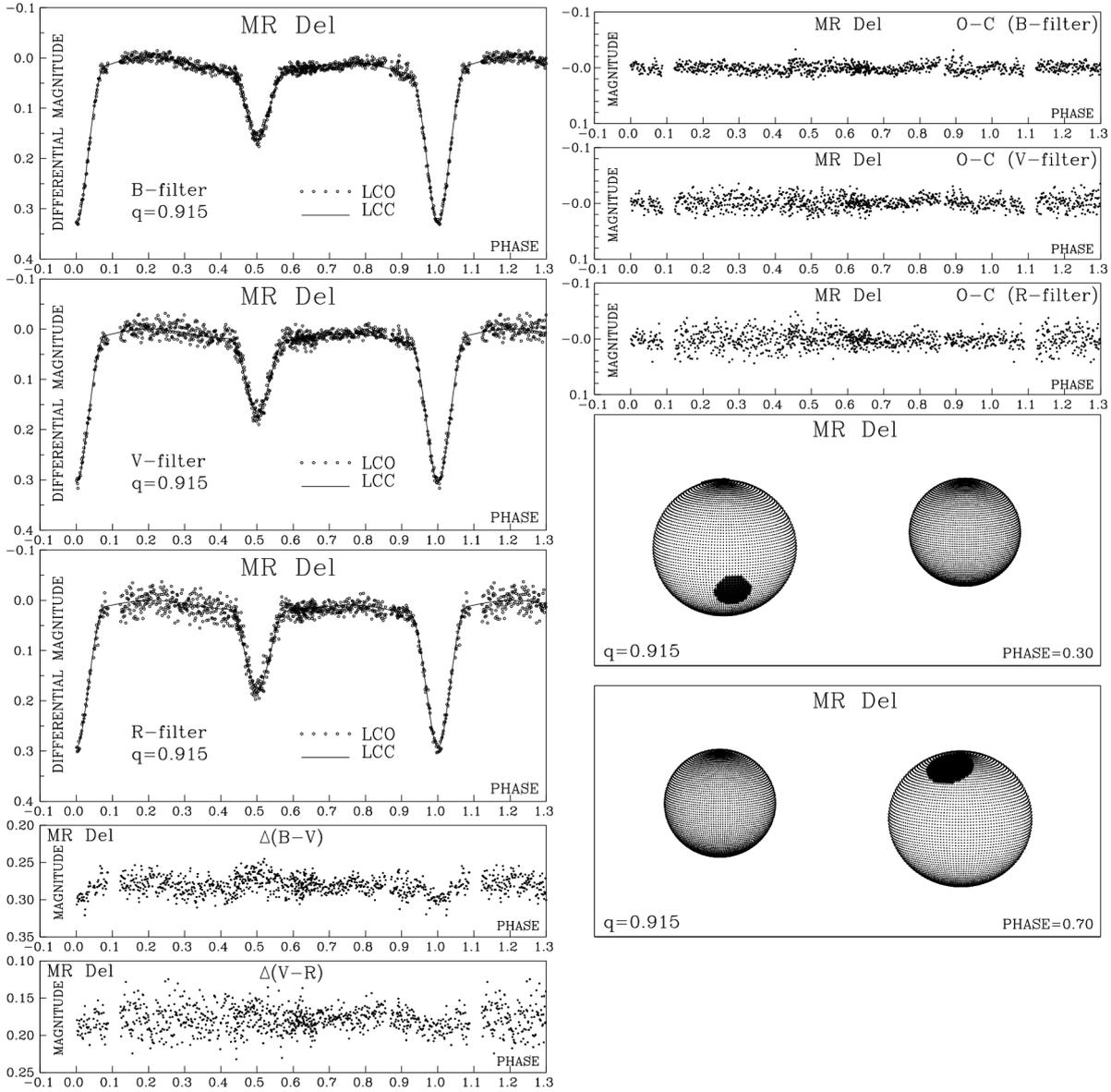}
\caption{Observed (LCO) and synthetic (LCC) light curves of {\rm MR Del},
obtained by simultaneous analysis of the B, V, and R observations,
$\Delta(B-V)$, $\Delta(V-R)$ color curve, final O-C residuals, and the
view of the system model at the orbital phases 0.30 and 0.70, obtained
with parameters from the inverse problem solution.}
\label{fMRDel}
\end{figure*}

An out-of-eclipse variation with an amplitude of 0.04 mag, observed in the V band light curve of the system, was interpreted as coming from a large spot covering 23\%  of the primary star's surface \citep{cuti97}. Although the system also displayed variations in X-ray emission, which were thought to be coming from the eclipsing component of the visual binary, it was not correlated with the light curve \citep{cuti97}.


\begin{table}[ht]
\begin{flushleft}
\caption{Results of the simultaneous (BVR) analysis of the {\rm MR Del}
light curves obtained by solving the inverse problem for the Roche model
with two cool spots on the less massive (hotter) component.}
\label{TabMRDel}
\[
\begin{array}{ll}
\hline
\noalign{\smallskip}
{\rm Quantity} &  \\
\noalign{\smallskip}
\hline
\noalign{\smallskip}
n & 2574 \\
{\rm \Sigma(O-C)^2} & 0.3120 \\
{\rm \sigma} & 0.0110 \\
q=m_{\rm h}/m_{\rm c} & 0.915 \pm 0.012 \\
T_{\rm h} & 4900 \\
A_{\rm c}=A_{\rm h} & 0.5 \\
\beta_{\rm c}=\beta_{\rm h}  & 0.08 \\
f_{\rm c}=f_{\rm h} & 1.0 \\
{\rm A_{S1}=T_{S1}/T_h} & 0.70 \pm 0.03 \\
{\rm \theta_{S1}} & 19.7 \pm 2.2 \\
{\rm \lambda_{S1}} &272.1 \pm 1.2 \\
{\rm \varphi_{S1}} & 61.5 \pm 5.7 \\
{\rm A_{S2}=T_{S2}/T_h} & 0.70 \pm 0.03 \\
{\rm \theta_{S2}} & 14.8 \pm 0.7 \\
{\rm \lambda_{S2}} & 100.1 \pm 1.3 \\
{\rm \varphi_{S2}} &-27.2 \pm 3.8 \\
T_{\rm c} &4400  \pm 20 \\
F_{\rm h} &0.74 \pm 0.01 \\
F_{\rm c} &0.62 \pm 0.01 \\
i \ [^\circ] &78.7  \pm 0.1 \\
{\rm L_3/(L_h+L_c+L_3) (B)}  &0.23 \pm 0.03 \\
{\rm L_3/(L_h+L_c+L_3) (V)}  &0.25 \pm 0.03 \\
{\rm L_3/(L_h+L_c+L_3) (R)}  &0.25 \pm 0.02 \\
\Omega_{\rm h} & 4.58 \pm 0.09 \\
\Omega_{\rm c} & 5.29 \pm 0.06 \\
R_{\rm h} [D=1] & 0.27 \pm 0.01 \\
R_{\rm c} [D=1] & 0.21 \pm 0.01 \\
{\rm L_h/(L_h+L_c+L_3)} & 0.59 [{\rm B}] ;\ 0.56 [{\rm V}] ;\ 0.56 [{\rm R}] \\
\cal M_{\rm c} {\rm [M_{\odot}]}  & 0.69 \pm 0.07 \\
\cal M_{\rm h} {\rm [M_{\odot}]}  & 0.63 \pm 0.06 \\
\cal R_{\rm c} {\rm [R_{\odot}]}  & 0.83 \pm 0.02 \\
\cal R_{\rm h} {\rm [R_{\odot}]}  & 0.65 \pm 0.02 \\
{\rm log} \ g_{\rm c} & 4.44 \pm 0.02 \\
{\rm log} \ g_{\rm h} & 4.61 \pm 0.02 \\
M^{\rm c}_{\rm bol} & 5.92 \pm 0.08 \\
M^{\rm h}_{\rm bol} & 6.92 \pm 0.08 \\
a_{\rm orb} {\rm [R_{\odot}]} & 2.99 \pm 0.03 \\
d[\rm pc] & 63 \pm 9 \\
\noalign{\smallskip}
\hline
\end{array}
\]
\end{flushleft}
\end{table}

The eclipsing component of the visual binary was later named MR Del by \citet{kaz99}. The dispute over the magnitudes of each of the components of the visual binary was ended by \citet{mason01}, who gave 9$^{m}$.49  for the V band magnitude of MR Del. \citet{clau01} obtained ubvy light curves of the system, and observed a flare event lasting for 25 minutes. They also point out the night-to-night differences in the light curves, increasing in strength from y to u band, which they attribute to surface activity.

\citet{pribb09} observed the system spectroscopically. They computed the light contribution of the visual companion as $\rm L_3/(L_1+L_2)=0.51$. The broadening functions they employed to obtain radial velocities did not show any evidence of photospheric spots. They found the mass ratio of 0.915 and emphasized the need for a new photometric analysis of the system to attain the absolute physical parameters.

\citet{zasche09} updated the light elements after having analyzed all photometric and astrometric data available for the system.  According to the observational indicators, MR Del has properties similar to stars of BY Dra type or of short-period RS CVn stars.

The results of our photometric analysis, based on updated spectroscopic elements of \citet{pribb09}, are given in Table~\ref{TabMRDel}. Figure~\ref{fMRDel} shows the observed (LCO) and the synthetic (LCC) light curves in the B, V, and R filters (upper left), the $B-V$ and $V-B$ color indices (lower left), the $O-C$ residuals (upper right) and the geometrical model of the system in representative phases 0.3 and 0.7 (lower right). Table~\ref{TabMRDel} lists parameter uncertainties estimated by combining the formal nonlinear least-squared fitting errors with the errors arising from the uncertainty of the spectroscopic mass ratio ($q=0.915 \pm 0.012$), as described in Section~\ref{analysis}.

Our model includes two cool spots on the more-massive, hotter component. The spotted model is supported by the X-ray observations. Another activity indicator is the flare event observed by \citet{clau01} which was most pronounced in the u band. In addition, there are night-to-night differences in the light curves, increasing in strength from the y to the u band, so cool spots can be expected on one or both components; however, the uniqueness of the spot locations obtained in our solution is questionable to some degree. A good fit could not be obtained with the spots located on the less massive secondary. A model with spots on the primary fits the observations very well, but fits of similar quality can be obtained with different spot locations, and we could not  decide definitely which of the possible solutions is the most appropriate.

Having in mind the vicinity of the third visual component, which cannot be resolved spatially,  we included a third light in the model. The estimated third light contribution to the total light of the system (${\rm L_3/(L_h+L_c+L_3)}\approx 0.23$) is significant, but somewhat lower than the contribution reported by \citet{pribb09} ($\rm L_3/(L_1+L_2)=0.51$, which equals ${\rm L_3/(L_h+L_c+L_3)}\approx 0.33$ for the luminosities of the eclipsing components in our model). There is also a considerable difference between our estimation of the distance of MR Del ($63 \pm 9 \rm pc$) and the  one computed based on the new Hipparcos parallax ($49 \pm 6 \rm pc$). The greater distance implies that the total luminosity of the system is overestimated in our model, which may be a result of complications introduced by the third light. We made trial runs with the $\rm L_3$ value fixed to the spectroscopic estimation of \citet{pribb09} and obtained plausible solutions, but the fits to the observations were significantly worse.

In comparison to the results that \citet{cuti97} obtained from the  analysis of their UBV(RI) and soft X-ray light curves, our study gives somewhat lower masses of the components (${\cal M}_{\rm c,\ Cutispoto}=0.64\ M_\odot$, ${\cal M}_{\rm h,\ Cutispoto}=0.72\ M_\odot$ and ${\cal M}_{\rm c,\ this\ study}=0.69\ M_\odot$, ${\cal M}_{\rm h,\ this\ study}=0.63\ M_\odot$), different radii (${\cal R}_{\rm c,\ Cutispoto}=0.70\ R_\odot$, ${\cal R}_{\rm h,\ Cutispoto}=0.77\ R_\odot$ and ${\cal R}_{\rm c,\ this\ study}=0.83\ R_\odot$, ${\cal R}_{\rm h,\ this\ study}=0.65\ R_\odot$), and a slightly lower inclination ($i_{\rm Cutispoto}\approx80^\circ$ and $i_{\rm this\ study}\approx79^\circ$). We note that their analysis was only based on photometric data.


\section{BD +07$^{o}$ 3142}

BD +07$^{o}$ 3142 (GSC 00380-00247, ASAS 162003+0707.4, PPM 162491)) was found to be a variable star by \citet{poj02} during their analyses of the ASAS data. Since then, only one time of primary minimum has been published by \citet{par07}.


\begin{figure*}
\includegraphics{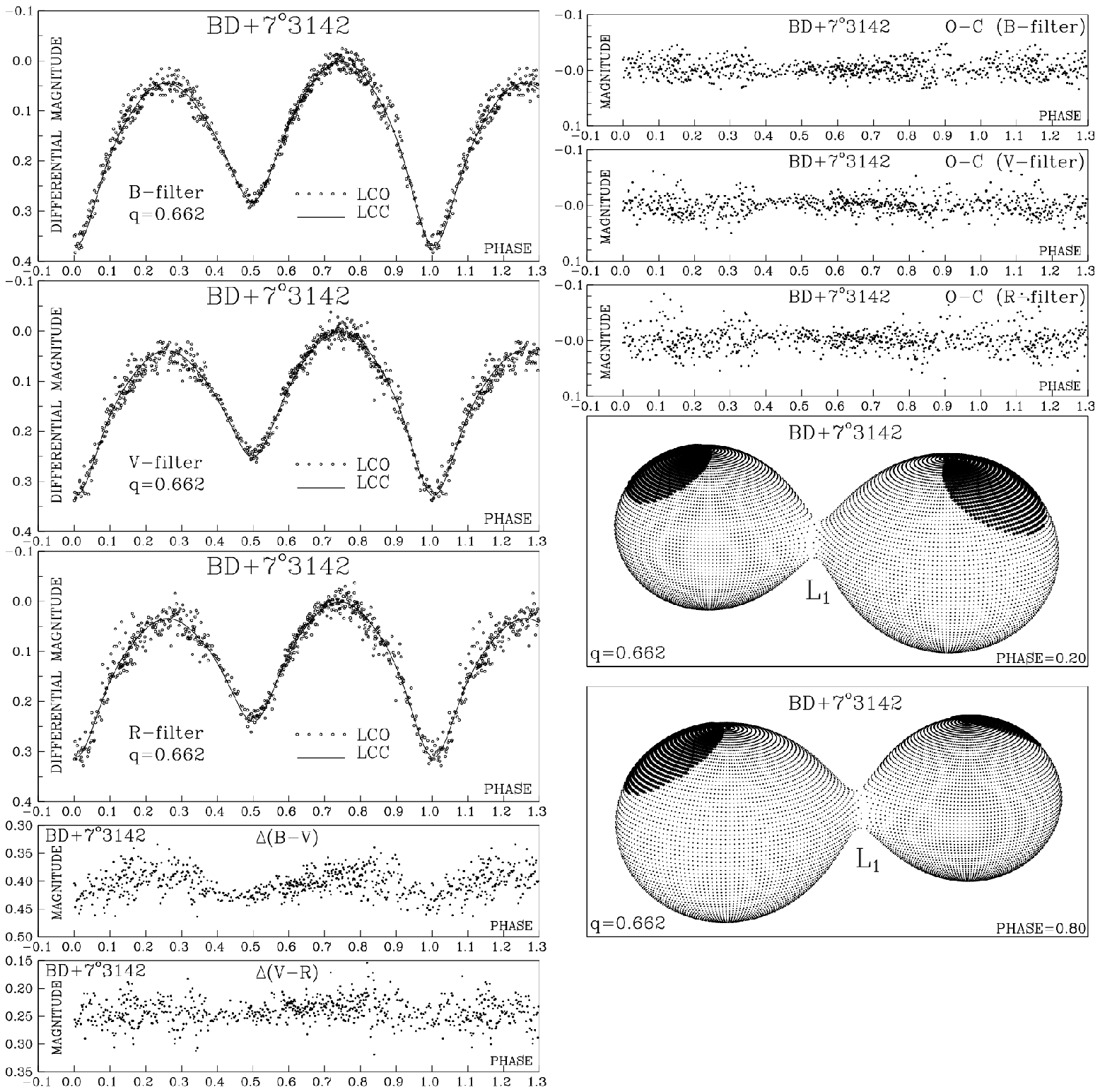}
\caption{Observed (LCO) and synthetic (LCC) light curves of ${\rm BD+7^o3142}$,
obtained by simultaneous analysis of the B, V, and R observations,
$\Delta(B-V)$, $\Delta(V-R)$ color curve, final O-C residuals, and the
view of the system model at the orbital phases 0.20 and 0.80, obtained
with parameters from the inverse problem solution.}
\label{fBD}
\end{figure*}

\citet{ruca08} observed the star spectroscopically to determine its radial velocity variation. They found a third component with a light contribution of L$_{3}$/(L$_{1}$ + L$_{2}$) = 0.50. Because of the proximity of the radial velocity of the third component to that of the center of mass of the binary system, they argue that it appears to be physically attached to the eclipsing pair. From the analysis of broadening functions, they determined the mass ratio of 0.662 and spectral type as K2V. They classified the system as a W UMa type binary.


\begin{table}[ht]
\begin{flushleft}
\caption{Results of the simultaneous (BVR) analysis of the ${\rm BD+7^o3142}$
light curves obtained by solving the inverse problem for the Roche model
with two cool spots, one on each component.}
\label{TabBD}
\[
\begin{array}{ll}
\hline
\noalign{\smallskip}
{\rm Quantity} &  \\
\noalign{\smallskip}
\hline
\noalign{\smallskip}
n & 1719 \\
{\rm \Sigma(O-C)^2} & 0.4572 \\
{\rm \sigma} & 0.0163 \\
q=m_{\rm h}/m_{\rm c} & 0.662 \pm 0.008 \\
T_{\rm h} & 4900 \\
A_{\rm h}=A_{\rm c} & 0.5 \\
\beta_{\rm h}=\beta_{\rm c}  & 0.08 \\
f_{\rm h}=f_{\rm c} & 1.0 \\
{\rm A_{s1}=T_{s1}/T_c} & 0.70 \pm 0.02 \\
{\rm \theta_{s1}} & 33.3 \pm 0.4 \\
{\rm \lambda_{s1}} &190.4 \pm 0.9 \\
{\rm \varphi_{s1}} & 53.8 \pm 0.7 \\
{\rm A_{s2}=T_{s2}/T_h} & 0.81 \pm 0.02 \\
{\rm \theta_{s2}} & 33.1 \pm 0.5 \\
{\rm \lambda_{s2}} &326.7 \pm 1.8 \\
{\rm \varphi_{s2}} & 56.9 \pm 1.0 \\
T_{\rm c} & 4640 \pm 30 \\
F_{\rm c} &1.014 \pm 0.002 \\
i \ [^\circ] & 71.6 \pm 0.3 \\
{\rm L_3/(L_h+L_c+L_3) (B)}  &0.451 \pm 0.004 \\
{\rm L_3/(L_h+L_c+L_3) (V)}  &0.478 \pm 0.003 \\
{\rm L_3/(L_h+L_c+L_3) (R)}  &0.480 \pm 0.004 \\
\Omega_{\rm h,c} & 3.140 \pm 0.015 \\
\Omega_{\rm in},\Omega_{\rm out} & 3.18 \pm 0.02 \ , 2.79 \pm 0.01 \\
f_{\rm over} [\%] & 9.5 \pm 1.3\\
R_{\rm c} [D=1] & 0.396 \pm 0.002\\
R_{\rm h} [D=1] & 0.328 \pm 0.002\\
{\rm L_h/(L_h+L_c+L_3)} & 0.274 [{\rm B}] ;\ 0.265 [{\rm V}];\ 0.268 [{\rm R}] \\
\cal M_{\rm c} {\rm [M_{\odot}]}  & 0.74 \pm 0.05 \\
\cal M_{\rm h} {\rm [M_{\odot}]}  & 0.49 \pm 0.04 \\
\cal R_{\rm c} {\rm [R_{\odot}]}  & 0.81 \pm 0.01 \\
\cal R_{\rm h} {\rm [R_{\odot}]}  & 0.67 \pm 0.01 \\
{\rm log} \ g_{\rm c} & 4.49 \pm 0.03 \\
{\rm log} \ g_{\rm h} & 4.48 \pm 0.03 \\
M^{\rm c}_{\rm bol} & 6.20 \pm 0.09 \\
M^{\rm h}_{\rm bol} & 6.37 \pm 0.09 \\
a_{\rm orb} {\rm [R_{\odot}]} & 1.91 \pm 0.02 \\
d[\rm pc] & 76 \pm 6 \\
\noalign{\smallskip}
\hline
\end{array}
\]
\end{flushleft}
\end{table}

The results of our photometric analysis, based on spectroscopic elements of \citet{ruca08}, are given in Table~\ref{TabBD}. This is the first ever published photometric solution of this system's light curves. Figure~\ref{fBD} shows the observed (LCO) and the synthetic (LCC) light curves in the B, V, and R filters (upper left), the $B-V$ and $V-B$ color indices (lower left), the $O-C$ residuals (upper right) and the geometrical model of the system in representative phases 0.2 and 0.8 (lower right). As in the previous sections, the uncertainties given in Table~\ref{TabQXAnd} were estimated by combining the formal nonlinear least-squared fitting errors with the errors arising from the uncertainty of the spectroscopic mass ratio ($q=0.662 \pm 0.008$), as described in Section~\ref{analysis}.

According to our analysis, ${\rm BD+7^o3142}$ is an overcontact binary ($f_{\rm over} \approx 10\% $)  in which the hotter, less massive component is eclipsed in the deeper primary minimum, so the system belongs to the W subclass of W UMa type binaries. The light curve asymmetry can be explained by the presence of two cool spots, located near the polar regions of each component.

Since the orbital period of the system is short  ($P\approx 0^d.28$), its components are in  fast rotation. Owing to the tidal effects, we expect a synchronization of the star rotation with the orbital period. According to \citet{Schusler}, the presence of spots at high latitudes in such a situation can be explained by the dynamo mechanism for rapid rotators.

The inclination of the orbit was estimated to be $i\approx72^\circ$, which means that both the primary and the secondary minima are due to partial eclipses. The third component (probably physically attached to the eclipsing pair) contributes approximately 48\% to the total light. Such a strong third light effect has a serious impact on the estimation of the basic system parameters and is particularly important for estimating the degree of overcontact and the orbital inclination.


\section{Conclusions}
We estimated the physical parameters of four binary systems -- QX And, RW Com, MR Del, and ${\rm BD+7^o3142}$ - from the analysis of new, high quality CCD light curves in the B, V, and R filters, and from the results of recent high-resolution spectroscopic studies. This is the summary of our findings.
   \begin{enumerate}
	\item{QX And is a close binary belonging to the A subclass of W UMa type systems, in a deep overcontact configuration with the degree of overcontact of $f\approx35\%$ and almost equal temperatures of the components ($T_h=6440 K$ and $T_c=6422 K$). The inclination of the orbit is estimated to be $i=54.6$ degrees, and the distance of the system is $d=416$ pc.} The best-fitting model contains a bright spot in the neck region of the hotter (more massive) component. 
	\item{RW Com is a close binary of W UMa type and W subtype in a shallow overcontact configuration with the degree of overcontact of $f\approx6\%$ and almost equal temperatures of the components ($T_h=4900 K$ and $T_c=4719 K$). The inclination of the orbit is estimated to be $i=74.9$ degrees, and the distance of the system is $d=91$ pc. The best-fitting model contains a bright spot in the neck region of the hotter (less massive) component.}
	\item{MR Del is a detached close binary and a component of a visual pair. Its complicated light curve was modeled with two cool spots on the hotter component. The estimated temperatures of the components are $T_h=4900 K$ and $T_c=4396 K$, and the inclination of the system is $i=78.4$ degrees. Our estimation of the distance of the system is $d=63$ pc. We find that the contribution of the third light in V band is $L_3/(L_h+L_c+L_3)=0.23$.}
	\item{${\rm BD+7^o3142}$ is a close, interacting binary. This is the first published analysis of its light curves. We find that it is an overcontact binary with a degree of overcontact of $f\approx10\%$, and we classify it as a W type W UMa system. The estimated temperatures of the components are $T_h=4900 K$ and $T_c=4645 K$ and the inclination of the system is $i=71.8$ degrees, with a considerable contribution of the third light in the V band of $L_3/(L_h+L_c+L_3)=0.48$. We estimate that the distance of the system is $76$ pc. The best-fitting model contains two large cool spots, one on each of the components. }
   \end{enumerate}
Since all four binaries appear to display surface activity, long-term observational campaigns along with the application of imaging techniques to affirm the existence and fix the locations of stellar spots would make valuable contributions to our knowledge about these systems. The question of whether ${\rm BD+7^o3142}$ belongs to a wide, visual system remains open and calls for further study.

\begin{acknowledgements}
The authors would like to thank T. Tanr{\i}verdi, A. Elmasl{\i}, and H.V. \c{S}enav{\i} for their help during the observations and
S. O. Selam, F. Ekmek\c{c}i and B. Albayrak for their suggestions on the paper. This research was funded in part by the Ministry of Science and Technological Development of the Republic of Serbia through the project ``Stellar and Solar Physics'' (No. 146003).  \"{O}. Ba\c{s}t\"{u}rk  would like to thank The Scientific and Technological Research Council of Turkey (T\"{U}BITAK) for the support by BIDEB-2211 scholarship. The authors acknowledge the use of the Simbad database, operated at the CDS, Strasbourg, France, and of NASA's  Astrophysics Data System Bibliographic Services.
\end{acknowledgements}

\end{document}